\documentclass[aps,prb,twocolumn,superscriptaddress]{revtex4-1}

\usepackage{graphicx}
\usepackage{dcolumn}
\usepackage{bm}
\usepackage{amsmath}
\usepackage{cases}

\usepackage{color}

\newcommand{\He}{$^4$He}
\newcommand{\et}{$\it{et \textrm{ } al. }$ }
\newcommand{\ie}{$\it{i. e.}$}
\newcommand{\dd}{\textrm{d}}
\newcommand{\B}[1]{\bm{#1}}

\begin{document}


\title{Spherically symmetric formation of localized vortex tangle around a heat source in superfluid $^4$He}



\author{Sosuke Inui}
\affiliation{Department of Physics, Osaka City University, 3-3-138 Sugimoto, 558-8585 Osaka, Japan}
\author{Makoto Tsubota}
\affiliation{Department of Physics \& Nambu Yoichiro Institute of Theoretical and Experimental Physics (NITEP) \& The OCU Advanced Research Institute for Natural Science and Technology (OCARINA), Osaka City University, 3-3-138 Sugimoto, 558-8585 Osaka, Japan}

\date{\today}

\begin{abstract}
We study the dynamical process of the vortex tangle development under a spherically symmetric thermal counterflow around a heat source submerged into a bulk superfluid \He.
We reveal a peculiar vortex dynamics that is unique to this geometry, which is greatly diverse from the vortex dynamics in a homogeneous counterflow.
Two types of heater are considered here, namely, a spherical heater with a solid wall and a point-like heater.
In both cases, a spherical vortex tangle is formed surrounding the heater.
The mechanism of vortex tangle development in the vicinity of a solid wall is strongly governed by Donnelly-Glaberson instability; while, far away from the heater or around a point heater, the mechanism is governed by the dynamics of polarized vortex loops in radial counterflow.
The decay process of such localized vortex tangles is also investigated and is compared with that of homogeneous vortex tangles.
\end{abstract}

\pacs{xxxx}

\maketitle


\section{Introduction}
%
The superfluid $^4$He has an extremely high thermal conductivity and acts as an excellent coolant in experiments conducted at extremely low temperatures \cite{VanSciver, Sciacca2015}.
The high thermal conductivity reflects the two-fluid nature of the superfluid $^4$He.
At $0$ K, the macroscopic quantum effect governs the entire fluid and forms a ground state without any entropy.
The type of flows allowed in such a system is either a potential flow or a circulatory flow around a filamentary topological defect with a quantized circulation, \ie a quantized vortex. 
On the other hand, at a finite temperature, the thermal excitations in the system form a viscous fluid with nonzero entropy $s$, resulting in a so-called normal fluid \cite{Tilley90, Donnelly91}.
%

%
The highly effective thermal transport is achieved by the normal fluid with velocity $\B{v}_n$ proportional to the heat flux $ \B{q}$ from a heat source per unit area, expressed as:
\begin{equation}\label{eq: normal - heat flux}
 \B{q} = \rho\B{v}_{n} \sigma T,
\end{equation}
where $\rho$ is the density of the fluid and $\sigma \equiv s/\rho V$ is the specific entropy that is dependent on the temperature $T$.
We consider a long closed pipe of cross-sectional area $A$ with a heater of heating power $W$ installed near one end and filled up with superfluid $^4$He. The normal-fluid component is driven from the hot end to the cold end, while the superfluid component is driven in the opposite direction to conserve the total mass of the fluid in the pipe.
The relative velocity $v_{ns} = | \B{v}_{n} - \B{v}_{s} | $ between the two fluids is thus given by,
\begin{equation}\label{eq: relative - heat flux}
 v_{ns} = \frac{1}{\rho_s \sigma T} \frac{W}{A},
\end{equation}
where $\rho_s$ is the density of the superfluid component.
While the heating power $W$ is smaller than some critical value $W_c$, the relative velocity $v_{ns}$ increases linearly with $W$.
A series of milestone experiments pioneered by Vinen, in the  late 1950s \cite{Vinen57_1,Vinen57_2,Vinen57_3,Vinen57_4} showed that the heating power dependence is significantly modified above the critical power $W_c$ \cite{Brewer,Chase,Childers,Tough82}.
It was concluded that the modification results from the formation of a quantized vortex tangle, which is identified as a turbulent state of the superfluid component or quantum turbulence \cite{Vinen02,Tsubota09,Tsubota13,Barenghi14}.
Since the core of a quantized vortex of radius $a \sim 1$ \AA{ } scatters the thermal excitations that constitute the normal-fluid component, a vortex filament effectively feels the mutual friction acting on it, which allows the energy transport between the normal fluid and the vortices.
The kinetic energy stored in a vortex filament per unit length is roughly estimated as $\varepsilon \sim \rho_s \kappa^2 /4\pi$ with quantized circulation $\kappa = h/m_\textrm{He}\approx 10^{-3}$ cm$^2$/s, where $h$ is Planck's constant and $m_\textrm{He}$ is the mass of a $^4$He atom.
This indicates that the energy transfer to the vortices results in the increase in the total length or/and the length density of a vortex tangle.
With a fully developed vortex tangle, where the vortex growth rate and the decay rate are balanced, the effective heat transport mediated by the normal-fluid is disturbed, and its effective thermal conductivity drops significantly.
%

%
The sudden drop in the thermal conductivity in the vicinity of a hot spot created in a superfluid $^4$He is a serious challenge in experiments because it may quench the entire system and prevent the fluid from working as an efficient coolant.
It is, therefore, important to understand how a vortex tangle evolves near a heater under a nonuniform thermal counterflow profile.
Furthermore, understanding the dynamical process of vortex tangle propagation under such a counterflow would provide insights on the ``lost energy'' in the micro big-bang experiment in Grenoble \cite{Bauerle, Bunkov}, as well as the ``peculiar motions'' of micron-sized particles trapped on the superfluid $^4$He surface \cite{Moroshkin19,Inui18}. 
The majority of preceding experimental\cite{Guo09, Guo10, Skrbek12, Marakov15, Svancara18, Hrubcova18} and numerical\cite{Schwarz85, Schwarz88,  Adachi10, Baggaley13, Baggaley15, Khomenko15, Yui15, Khomenko17, Yui2018,Yui2020} studies mainly addressed the properties of steady vortex tangles in a thermal counterflow in a narrow channel .
However, the tangle formation processes in different thermal counterflow profiles have rarely been studied until the recent numerical works carried out by Varga \cite{Varga18} and Sergeev \et \cite{Sergeev19}.
In this paper, we present the results of our numerical simulation on the investigation of localized inhomogeneous vortex tangles formed around a spherical heat source immersed in the superfluid.
The decay process of such localized vortex tangles is also compared with the homogeneous ones in terms of the phenomenological parameter $\chi_2$ in the Vinen's equation \cite{Vinen57_3}.
%

%
We suppose a small number of remnant vortices in the system and simulate their time evolution based on the vortex filament model (VFM) under a spherically symmetric steady thermal counterflow.
In the simulations, vortices are allowed to reconnect to the spherical wall of the heater, so that a role of a solid boundary in the real experiments can be investigated.
Also, in this numerical study, we suppose a steady thermal counterflow, \ie, a fixed normal-fluid flow profile is prescribed, and disturbances in its profile are neglected.
The feedback on the normal-fluid profile due to the vortices through the mutual friction is not discussed as it is beyond the scope of this paper.
%

The paper is organized as follows:
In section \ref{sec: D of V}, we briefly introduce our numerical scheme and the overview of single vortex dynamics.
In Sec. III, we present the results of VFM simulations and discuss the dynamical processes of the tangle development around a spherically symmetric thermal counterflow. 
The heat source is assumed to have a solid wall to which vortices can connect, and we also present the analysis of the structure of the vortex tangle.
The heater discussed in Sec. IV is point-like, and we investigate the vortex evolution at some distance apart from the heat source. 
Further, we discuss the characteristics of the decay process of the localized vortex tangles.
Finally, the summary is presented in Sec. V.
 %
%
 
\section{Dynamics of Vortices} \label{sec: D of V}
\subsection{Numerical Method: Vortex Filament Model}
%
%
%
%

%
The motion of a quantized vortex follows the local superfluid flow, as described by Helmholtz's theorems.
Taking into account  the temperature-dependent mutual frictions, $\alpha$ and $\alpha^\prime$, the equation of motion of a vortex segment at $\B{s}(\xi)$ is \cite{Schwarz85}
\begin{equation} \label{eq: EoM VFM}
\begin{split}
\frac{\dd \B{s}(\xi,t)}{\dd t} = \B{v}_\textrm{s} & + \alpha \B{s}^\prime(\xi)  \times \B{v}_\textrm{ns} \\
&- \alpha^\prime \B{s}^\prime (\xi)  \times  \left[ \B{s}^\prime (\xi) \times \B{v}_\textrm{ns} \right] ,\\
\end{split}
\end{equation}
with
\begin{equation} \label{eq: v_s}
\B{v}_s = \B{v}_{s,\textrm{ind}} +    \B{v}_{s,\textrm{BS}} \quad \textrm{and} \quad \B{v}_n = \B{v}_{n,\textrm{ind}}.
\end{equation}
Here, $\xi$ is the arc-length parameterization of the vortex filaments, and $\B{s}^\prime(\xi)$ is the unit normal vector along the vortex filament at $\xi$, and $ \B{v}_{s,\textrm{ind}}$ and $ \B{v}_{n,\textrm{ind}}$ refer to the velocity fields thermally excited by the heater.
The superfluid velocity $\B{v}_{s,\textrm{BS}}$ includes velocity contribution from all the vortices, and is written in the form of Biot-Savart integral;
\begin{equation} \label{eq: BS integral}
\begin{split}
\B{v}_{s,\textrm{BS}} =& \frac{\kappa}{4\pi} \int_{\mathcal{L}} \frac{ \B{s}^{\prime} (\xi) \times ( \B{s} (\xi_0) - \B{s} (\xi) ) }{ | \B{s} (\xi_0) -\B{s}(\xi)|^3 } \dd \xi   \\
 =& \B{v}_{s,\textrm{loc}} + \B{v}_{s,\textrm{non-loc}}.  \\
\end{split}
\end{equation}
The divergence of the integral in the right-hand-side of Eq. \eqref{eq: BS integral} as $\xi \rightarrow \xi_0$ can be avoided by separating it into two terms, a localized induction velocity $\B{v}_{s,\textrm{loc}} \approx \beta \B{s}^\prime \times \B{s}^{\prime\prime}$ and a nonlocal contribution $\B{v}_{s,\textrm{non-loc}}$, where $\beta = (\kappa/4\pi) \ln(R/a)$ and $\B{s}^{\prime\prime}$ is the second derivative of $\B{s}(\xi)$ with respect to $\xi$.
The interaction between the normal fluid and the vortices takes place through $\B{v}_{ns} = \B{v}_{n} - \B{v}_{s}$ in the mutual friction terms in Eq. \eqref{eq: EoM VFM}. 
By discretizing the vortices in line segments of a suitable resolution $\Delta \xi$, we solve the integro-differential equation.
Also, the time-integration is calculated with a suitable time resolution $\Delta t$, adopting the fourth-order Runge-Kutta scheme.
%

%
To simulate a realistic spherical heater onto which vortices can be trapped, calculation of vortex dynamics is subjected to a spherical solid boundary condition.
The condition can be satisfied by finding a boundary induced velocity field $\B{v}_{s,b}$ such that
\begin{equation} \label{eq: BC}
\left( \B{v}_{s,BS} + \B{v}_{s,b} \right) \cdot \B{\hat{n}} = 0
\end{equation}
holds at the surface of the sphere of radius $r_0$.
Here, $\B{\hat{n}}$ is the unit normal vector on the surface.
Since the velocity $\B{v}_{s,b}$ satisfies the system of equations
\begin{equation} \label{eq: v_s boundary}
\begin{split}
\nabla \times \B{v}_{s,b} &= 0 \\
\nabla \cdot \B{v}_{s,b} &= 0, \\
\end{split}
\end{equation}
we can solve Eq.\eqref{eq: v_s boundary} in terms of the associated Legendre polynomials \cite{Schwarz85}.

\subsection{Motion of a Vortex Ring in Spherical Thermal Counterflow} \label{sec: D of V 2}
\begin{figure}[t!] 
	\includegraphics [width=1\columnwidth] {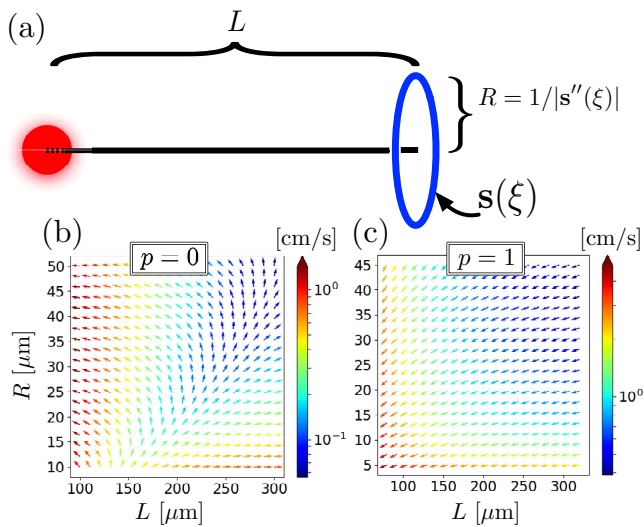}
	\caption{(a) Schematics of a spherical heater and a vortex ring of radius $R$ separated by a distance $L$. In the absence of the heater, the vortex travels in the direction of $\B{s}^{\prime} \times \B{s}^{\prime\prime}$.  The heater tends to ``pull'' the vortex toward it. (b) and (c) The vector field representation of Eq.\eqref{eq: EoM simple} at $T=1.6$ K with $p=0$ and $p=1$, respectively (color online). The color indicates the magnitude of the flow $\sqrt{\dot{L}^2 + \dot{R}^2}$ at each location.}
	\label{fig: RING_GEOM}
\end{figure}
%
For the sake of clarity in the later discussion, we shall consider a single vortex ring traveling toward and away from a point heater immersed into a superfluid \He{} bath at $T = 1.6$ K as shown in Fig.\ref{fig: RING_GEOM}(a).
The heater is a sphere of radius $a = 100$ $\mu$m, around which we prescribe the spherical thermal counterflow such that the relative velocity between normal fluid and superfluid at some distance $r$ from the center of the heater is $\B{v}_\textrm{ns}(\B{r}) = \frac{1}{\rho_\textrm{s} \sigma T} \frac{W}{4 \pi r^2}\hat{\B{r}}$.
 This can be obtained by replacing the area $A$ in Eq.\eqref{eq: relative - heat flux} with $4\pi r^2$.
We consider Eq.\eqref{eq: EoM VFM} for this case.
By ignoring the nonlocal term in Eq.\eqref{eq: BS integral}, the superfluid velocity $\B{v}_\textrm{s}$ in Eq.\eqref{eq: v_s} becomes
\begin{equation} \label{eq: v_s expand}
\B{v}_\textrm{s} \approx \beta \B{s}^{\prime} \times \B{s}^{\prime \prime} - \displaystyle\frac{\rho_\textrm{n}}{\rho_\textrm{s}}  \frac{v_0}{L^2 + R^2}\hat{\B{r}},
\end{equation}
where $v_0 \equiv W/4\pi \rho\sigma T$, and $L$ is the distance between the centers of the heater and the vortex ring.
Substituting Eq.\eqref{eq: v_s expand} into Eq.\eqref{eq: EoM VFM}, one obtains a system of differential equations for the distance $L$ and the vortex radius $R$
\begin{subequations}\label{eq: EoM simple}
  \begin{align}
    & \dot{L} \approx  (-1)^p\frac{\beta}{R} -  \frac{\rho_\textrm{n}}{\rho}\frac{v_0}{L^2 + R^2}\textrm{cos}\theta -  (-1)^p \frac{\alpha v_0}{L^2 + R^2}\textrm{sin}\theta \label{eq: EoM simple x} \\
    & \dot{R}\approx  - \frac{\rho_\textrm{n}}{\rho} \frac{v_0}{L^2 + R^2}\textrm{sin}\theta - \alpha \left[ \frac{\beta}{R} - (-1)^{p} \frac{ v_0}{L^2 + R^2}\textrm{cos}\theta \right],  \label{eq: EoM simple R}  
   \end{align}
\end{subequations}
where $\theta = \tan^{-1}(R/L)$ and the terms with $\alpha^\prime$ are neglected.
In Eq.\eqref{eq: EoM simple}, $p$ is a parameter indicating the direction of the ring propagation; $p=0$ if the orientation of the vortex, \ie{} $\B{s}^{\prime}\times \B{s}^{\prime\prime}$, is radially outward, and $p=1$ if that is radially inward.
Figures \ref{fig: RING_GEOM}(b) and (c) show Eq.\eqref{eq: EoM simple} as vector fields $ (  \dot{L}(L,R),\dot{R}(L,R) ) $ with $p=0$ and $p=1$, respectively, at $T = 1.6$ K.
The trajectory of a vortex can be obtained by drawing a smooth stream line parallel to the vector at each point from an arbitrary initial point $(L_0, R_0)$ in the coordinate.
In the case with $p=1$, a vortex of any radius $R$ shrinks itself to vanish at some distance $L$ away or ``collides with'' the heater at $L = 100$ $\mu$m with time.
On the other hand, when $p = 0$, a vortex could grow in size depending on its initial values, $R_0$ and $L_0$.
%

%
Though in a highly developed vortex tangle, this simple single-loop analysis may not hold without modification, we can learn from this analysis that the orientation of each vortex ring that constitutes the tangle tends to orient itself in the outward direction. 
In preceding studies with a simple channel flow geometry, the homogeneity or/and isotropy of the vortex orientations are often assumed, which allows us to derive the Vinen's equation which predicts the time-evolution of a vortex line density \cite{Vinen57_3}.
However, in the present study there is no guarantee that the same equation holds because the vortex orientations are polarized.
%
%

\section{Vortex Tangle near a Spherical Heater}
\subsection{Development of a Vortex Tangle}
\begin{figure*}[t!]
	\includegraphics [width=2\columnwidth]{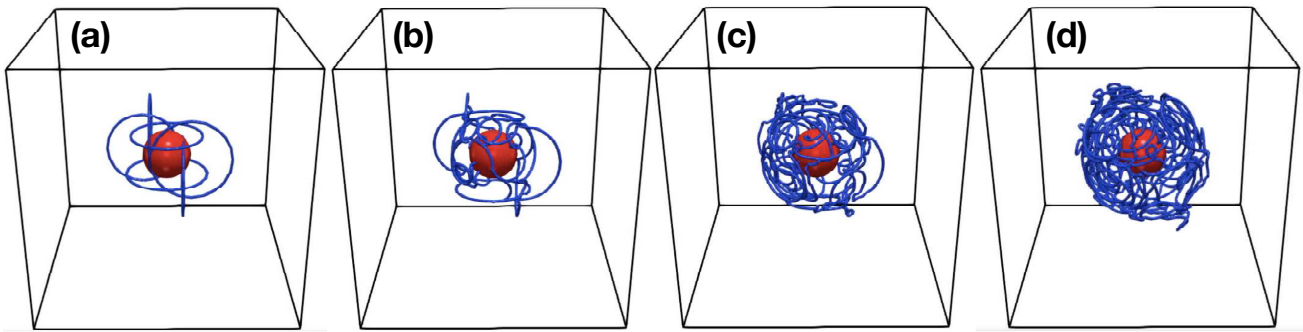}
	\caption{(a)-(d) Snapshots of development of a vortex tangle (blue filaments) around a heat source of radius $0.1$ mm (the red sphere) at $t = 0.0$, $2.0$, $4.0$, and $6.0$ ms, respectively. A cube of length $1$ mm is drawn as a reference in each panel. }
	\label{fig: Sphere time evol}
\end{figure*}
A sphere of a radius $r_0 = 0.1$ mm is immersed in a bulk superfluid \He.
The length of each vortex segment $\Delta \xi$ is set to be within the range $1 \sim 2 \times 10^{-2}$ mm. 
Now, the sphere is assumed to be heated homogeneously and sets up a steady thermal counterflow profile of the form
\begin{equation} \label{eq: Heater_Simulation}
\B{v}_{ns}  = v_{ns,0}\frac{r_0^2}{r^2}  \B{\hat{r}},
\end{equation}
where $v_{ns,0}$ is the magnitude of the relative counterflow velocity at the surface $r = r_0$.
In reality there should exist a ``thermal boundary layer'' with finite thickness where the two fluids are accelerated and the profile in Eq.\eqref{eq: Heater_Simulation} is not valid in the vicinity of the heater, as it is reported in Refs. \cite{Svancara18, Hrubcova18}. 
However, we ignore such effect in this study, assuming the width of the layer is much less than the particle radius $r_0$.

Figure \ref{fig: Sphere time evol} shows the time evolution of six seed vortices symmetrically placed near a heater.
%
%
\begin{figure}[t!] 
	\includegraphics [width=1\columnwidth]{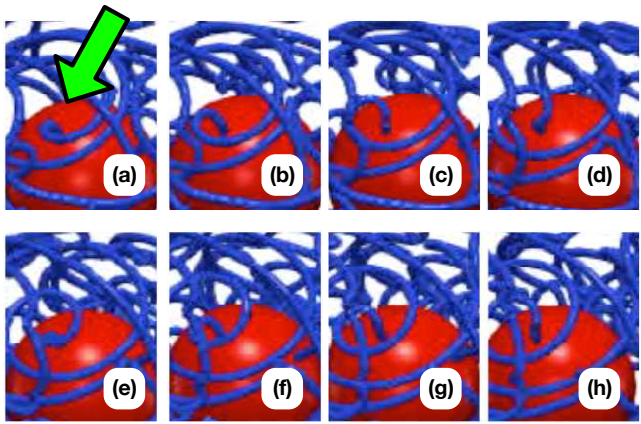}
	\caption{(a)-(h) Magnified snapshots of the development of the same vortex tangle.  They are taken every after $\Delta t = 0.2$ ms from $t = 4.0$ ms.  Spiral structures are likely to be excited on an edge of a vortex filament that reconnects the heater surface.}
	\label{fig: Sphere time evol 2}
\end{figure}
In the simulation, we set the temperature of the system to be $T = 1.3$ K and the counterflow velocity to be $v_{ns,0} = 100$ cm/s.
Out of the six vortices, four are initially reconnected onto the spherical surface.
An end of the vortex line reconnected to the heater surface is bent in order to meet the boundary condition of Eq.\eqref{eq: BC}, \ie, the vortex segment at the surface needs to be perpendicular to the surface so that the vertical velocity component vanishes where they meet on the surface.
As we discussed in Sec.\ref{sec: D of V 2}, under a radial counterflow, the orientation of vortices are selected in a way that they tend to induce velocity away from the heater, which seems to explain a spiral-shape structure formed near the vortex-surface intersection (see Fig.\ref{fig: Sphere time evol 2}).
The rapid growth of the tangle due to the spiral-shaped structure is also understood as the manifestation of the Donnelly-Glaberson (DG) instability \cite{Schwarz90,Glaberson66,Donnelly91}.
In the vicinity of the heater surface, the magnitude of the counterflow is strong, which makes the vortex lines unstable and excites Kelvin waves along them.
Again, the Kelvin waves with outward orientation are likely to be amplified; while the others tend to vanish because of the converging counterflow.
This growth mechanism driven by DG instability would not hold for vortices that traveled far enough from the heater because the counterflow magnitude drops as $1/r^2$.
We discuss the vortex tangle growth mechanism out of the DG instability regime in Sec. IV.
\subsection{Radial Vortex Line Density}
\begin{figure}[t!] 
	\includegraphics [width=1\columnwidth]{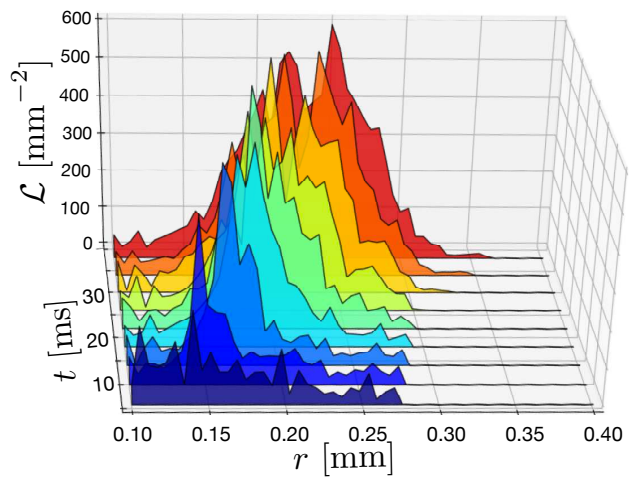}
	\caption{ Radial vortex line density (RVLD) $\mathcal{L}$ (color online). A single curve plots RVLD of the tangle at some time $t$.  $r$ is the radial distance from the center of the heater of radius $r_0 = 0.1$ mm.}
	\label{fig: Sphere time evol 3}
\end{figure}

The vortex line density of the tangle as a function of the radius is plotted in Fig.\ref{fig: Sphere time evol 3}.
A single curve in the plot shows the radial vortex line density (RVLD) at time $t$ in the entire tangle, and the change in color (blue $\rightarrow$ red) indicates the lapse of time (early $\rightarrow$ late).
Although the maximum RVLD has an upper bound due to the numerical limitation ($L_\textrm{max} \sim  \Delta \xi ^{-2}$), the tangle seems to grow unboundedly, as long as some vortices are kept reconnected onto the heater surface at $r=r_0 (= 0.1$ mm).
%

With time, a bottleneck appears in RVLD near the heater, which may be understood as follows:
The vortices in the tangle are constantly pulled toward the center of the heater by its induced superfluid velocity $\B{v}_{s,\textrm{ind}}$.
At the same time, a free vortex ring travels approximately with a localized induction velocity $\B{v}_{s,\textrm{loc}}$ in Eq.\eqref{eq: BS integral}, which is inversely proportional to the local curvature radius.
Therefore, a smaller radius of a vortex ring results in  a stronger resistance to the counter flow that pulls it toward the heater.
In the vicinity of the surface, only a few vortices can survive in the strong counterflow, while others are `sucked' into the heater.
On the other hand, the counterflow becomes weaker as $1/r^2$, and more and more vortices can overcome the pull as they go farther away from the heater.
%
%
\section{Vortex Tangle Around a Point Heater}
\begin{figure*}[t!] 
	\includegraphics [width=2\columnwidth]{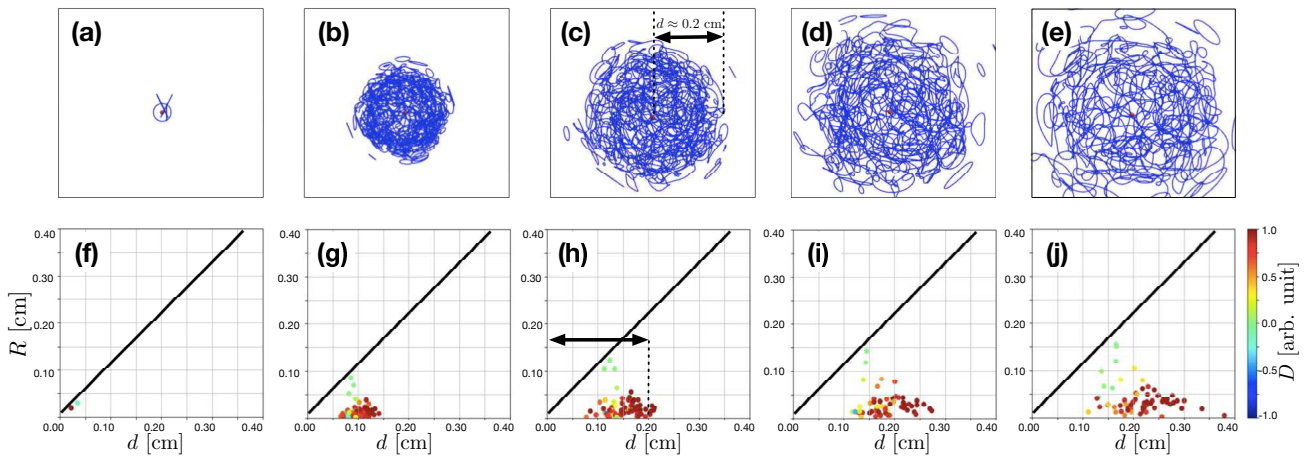}
	\caption{(a) - (e) Snapshots of vortices that grow to be a vortex tangle around the point heater. The parameters, $v_{ns}$ and $T$ are set to be $50$ cm/s and $1.9$ K, respectively. (f)--(j) Vortex size distribution in the tangle at corresponding time (color online). Each dot in the plots represents a vortex loop with some size $R$ (vertical axis) at some distance $d$ (horizontal axis) with some orientation indicated by the loop polarization $D$ (color from red to blue).  See Eq.\eqref{eq: Def D} and the text for the definition of $D$.}
	\label{fig: Point time evol}
\end{figure*}
In Sec. III we have considered the case where the initial seed vortices are placed relatively close to the heater, and some of them are reconnected onto it.
In this section, however, we treat the heater as a point in order to investigate the tangle development process far away from it, assuming no vortices are trapped on the heater.

\subsection{Development of a Tangle}  \label{sec: IV A}
%
Supposing a heater placed at the origin is point-like, we ignore the boundary condition of Eq.\eqref{eq: BC}, but the radius of the heater is set to be $r_0 = 0.1$ mm to obtain the counterflow velocity profile according to Eq.\eqref{eq: Heater_Simulation}.
The vortex resolution is $\Delta \xi \sim 0.5 \times 10^{-2}$ mm for this simulation.
By placing several seed vortices we investigated their development into a vortex tangle, and its dependence on the counterflow magnitude $v_{ns}$ and the temperature $T$.
%

A typical time-evolution process is summarized in Fig.\ref{fig: Point time evol}.
The panels (a)--(e) in Fig.\ref{fig: Point time evol} show the snapshots of the simulation.
A single dot in the panels (f)--(j) of Fig.\ref{fig: Point time evol} shows a vortex loop in the tangle corresponding to the moments when the snapshots (a)--(e) are taken.
The horizontal and vertical axes in the panels (f)--(j) represent the radial distance $d$ of a vortex loop, defined as the average distance between the origin and each vortex segment within a single loop, and the size $R$ of the loop is defined as the average distance between each vortex segment and the center of the loop. 
In terms of the lengths defined in Fig.\ref{fig: RING_GEOM}(a), the radius distance $d$ is expressed as $ \sqrt{ L^2 + R^2} $.
The color of the dot in the panels indicates the loop polarization $D$, which is defined as
\begin{equation} \label{eq: Def D}
D = \hat{\B{r}} \cdot  \oint_\textrm{a loop}  \B{s}^{\prime}(\xi) \times \B{s}^{\prime\prime}(\xi)\dd \xi / \mathcal{N},
\end{equation}
where $\hat{\B{r}}$ is the unit radial vector at the center of the loop, and $\mathcal{N}$ is a properly chosen numerical factor that normalizes the integral.
The vortex loops with $D\approx 1$ travel radially outward and are colored in red; while, the ones with $D\approx -1$ travel radially inward and are colored in blue.
It turns out that the small number of initial vortex loops grows swiftly.
The loops quickly form a spherical tangle surrounding the point heater.
The tangle front propagates radially outward, as it leaves a hollow region inside the tangle.
Solid slopes of roughly $d = R$ are drawn for reference in Fig.\ref{fig: Point time evol}(f)--(j).
Apparently, the maximum loop size $R_\textrm{max}$ appears in the vicinity of the slopes at each panel, which indicates that the value of $R_\textrm{max}$ can be used as a good measure of the radius of the tangle.
%

The existence of the large vortex loops of size comparable to the tangle size with polarization $D \approx 0$ seems to play the key role for the tangle front propagation.
Such large loops frequently collide with each other and emit small vortices with random orientations.
The orientations, however, are polarized because of the radial thermal counterflow.
As we have seen in Sec. II, small vortices with inward orientation shrink, while ones with outward orientation can survive for longer time.
Therefore, if the radii of outward vortices are small enough to overcome the counterflow that tries to suck them into the heater, then the radius of the vortex tangle can grow, which would explain the monotonous growth in radius of the tangle with a hollow inside.
This tangle front propagation process driven by small outward vortices can be observed at the outmost  region of the tangle in Fig.\ref{fig: Point time evol}(h), (i) and (j).
Also, see movies in supplementary materials.
%

The tangle front propagates radially outward against the superfluid velocity $\B{v}_{s,\textrm{ind}}$ induced by the point heater.
The relative counterflow profile  $\B{v}_{ns}$ at the tangle front $R_\textrm{f}$ is given by Eq.\eqref{eq: Heater_Simulation}, equating $r = R_\textrm{f}$.
Assuming that the conservation of mass is valid locally, the velocity $\B{v}_{s,\textrm{ind}}$ can be obtained from the following relation:
\begin{equation}
\rho_s \B{v}_{s,\textrm{ind}}(R_\textrm{f}) + \rho_n \B{v}_{n,\textrm{ind}}(R_\textrm{f}) = 0,
\end{equation}
where $\B{v}_{n,\textrm{ind}} - \B{v}_{s,\textrm{ind}} = \B{v}_{ns}$.
Since a vortex ring of radius $R$ travels with velocity $v_s \sim \beta/R$, the only vortex loops with radius $R_c < \beta/v_{s,\textrm{ind}}(R_\textrm{f})$ can travel against the counterflow and gradually expand the ``edge'' of the tangle, while the other vortices are presumably well-confined within the region of radius $R_\textrm{f}$.

\subsection{Total Vortex Line Length (TVLL) \\ and Decay Process} \label{sec: 4B}
%
\begin{figure}[b!] 
	\includegraphics [width=1\columnwidth]{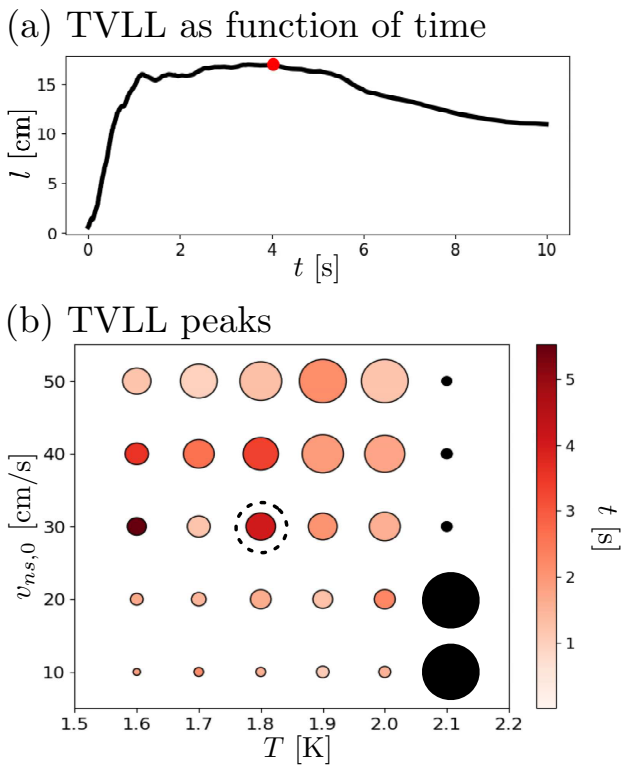}
	\caption{(a) Total vortex line length (TVLL) $l$ as function of time $t$ for $T = 1.8$ K and  $v_{ns,0} = 30$ cm/s.  The solid circle represents the maximum length $l_\textrm{max}$.  (b) TVLL peaks $l_\textrm{max}$ as function of $T$ and $v_{ns,0}$ (color online). The radii and the depth of the circle's face-color represent the maximum length $l_\textrm{max}$ and the time it takes to reach the length at each point in the parameter space $(T, v_{ns,0})$, respectively.  However, at $(T, v_{ns,0}) = (2.1, 20)$ and $(2.1, 10)$, $l(t)$ tends to grow unboundedly with time, so the radii do not reflect the actual vortex line length.}
	\label{fig: Point time evol 2}
\end{figure}
%
When it comes to the study of the homogeneous vortex tangles, the vortex line density (VLD) $\mathcal{L}$ is the quantity that is uniquely determined experimentally and computationally and is frequently used to analyze the nature of the tangles.
However, in the case of the localized inhomogeneous vortex tangles, there remains some arbitrariness in the value of $\mathcal{L}$ depending on the scheme to estimate the volume $V$ occupied by the vortex lines in the tangle.
The maximum value of spatially-independent VLD $\mathcal{L}$ may be achieved if the volume $V$ is estimated by the box-counting method, \ie, dicing up the computational space into cubes with suitable lengths, we count the number of the cubes that contain the (segments of ) vortices and calculate the total volume $V_\textrm{BC}$.
If we take the lengths of the cubes to be the average vortex loop size, then $\mathcal{L} = l/V_\textrm{BC}$, where $l$ is the total vortex line length (TVLL), gives the VLD averaged over the volume.
On the other hand, since most of the vortices in the tangle are localized in a sphere of volume $V_\textrm{loc} = 4\pi R_\textrm{f}^3 /3 $, the minimum value of $\mathcal{L}$ is achieved when we let the volume $V$ be a constant such that $V > V_\textrm{loc}$ for all time.
Then the TVLL $l$ in this case is essentially the same as the VLD $\mathcal{L}$, which is the quantity we shall mainly discuss in this section.
%

Figure \ref{fig: Point time evol 2}(a) shows the profile of the TVLL $l$ as a function of time.
Initially, $l(t)$ grows rapidly within the region of radius $R_\textrm{f}$.
As we have discussed in Sec. \ref{sec: IV A}, in the initial growing stage, there are number of large vortex loops of the order of the tangle size.
Since a large vortex loop cannot travel against the thermal counterflow, those with large radii or small local curvatures are essentially enclosed in the spherical region, inside of which vortices grow in length via mutual friction and in number because of the repeated reconnections.
However, the growth rate decreases, and the total length $l(t)$ finds its maximum value $l_\textrm{max}$.
In the case of $T = 1.8$ K and $v_{ns,0} = 30$ cm/s, the tangle grows up to $l_\textrm{max} \approx 15$ cm at time $t \approx 4$ s, after having some fluctuation around the maximum value for about a few seconds.
 Then, it finally starts to decay gradually.
 The dependence of $l_\textrm{max}$ on $T $ and $v_{ns,0} $, is summarized in Fig.\ref{fig: Point time evol 2}(b).
 The radii of the circles are proportional to the values of $l_\textrm{max}$ at $(T , v_{ns,0} )$, and the thickness of the circle color indicates the time it takes to reach $l_\textrm{max}$.
 Below $T \approx 2.0$ K, the TVLL profile $l(t)$ has a plateau for several seconds after a swift growth, during which the maximum value $l_\textrm{max}$ is attained. 
 Above $T \gtrsim 2.1$ K, on the other hand, the tangle behaves differently since the value of the mutual friction coefficient $\alpha^\prime$ becomes negative.
 If $T \gtrsim 2.1$ K, there is a critical velocity $v_c$, between $v_{ns,0} = 20$ cm/s and $30$ cm/s, that determines whether a tangle can develop or not.
 Above the critical velocity $v_c$, the tangle may not be formed; while, below $v_c$,  the tangle tends to grow unboundedly.
 %

\begin{figure}[b!] 
	\includegraphics [width=1\columnwidth]{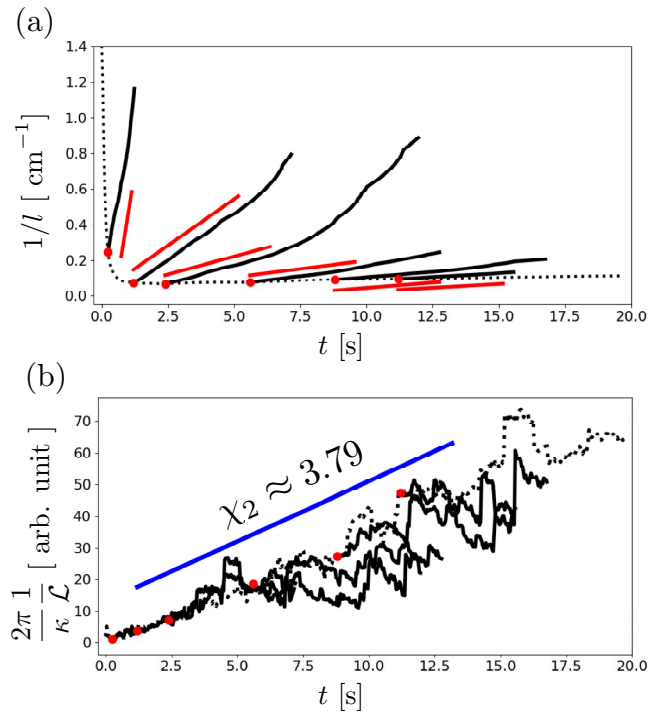}
	\caption{ (a) Comparison between ``gradual decay'' and ``free decay'' (color online ). The system is kept at $T = 1.8$ K, and the relative velocity before turning of the heater is $v_{ns,0} = 30$ cm/s. (b) Normalized $1/l$ plots for both ``gradual decay'' and ``free decay''.  The TVLL $l$ is normalized by the volume $V_\textrm{BC}$.}
	\label{fig: Point decay}
\end{figure}

Aside from the ``gradual decay'' while the heater is on, we have also investigated the ``free decay'' turning off the heater suddenly during the tangle development process.
Figure \ref{fig: Point decay}(a) plots $1/l$ as a function of time $t$. 
The broken curve represents the``gradual decay" where the heater is kept on.
At time $t_\textrm{free}$ indicated by the solid circles on the $1/l$ plot, the point heater is turned off, and the time evolutions of the tangle for time $t > t_\textrm{free}$ are simulated.
The solid branches departing from the broken curve represent the results of the  ``free decay'' simulations.
%
%
After turning off the heater, the TVLL $l$ tends to decay inversely proportional to time for several seconds as it is indicated by the straight lines in Fig. \ref{fig: Point decay}(a).
What this indicates is that the form of the decay term coincides with that of Vinen's equation (VE) \cite{Vinen57_3}
\begin{equation} \label{eq: VE decay}
\left(\frac{\dd \mathcal{L}}{\dd t} \right)_\textrm{decay} = - \chi_2 \frac{\kappa}{2 \pi}  \mathcal{L}^2
\end{equation}
that describes the time evolution of a homogeneous VLD $\mathcal{L}$, where $\chi_2$ is a phenomenological parameter of order unity.
Substituting $\mathcal{L} = l/V_\textrm{BC}$ into Eq.\eqref{eq: VE decay}, we obtain
\begin{equation} \label{eq: Decay no-approx}
\frac{\dd l}{\dd t} - \frac{l}{V_\textrm{BC}}\frac{\dd V_\textrm{BC}}{\dd t}= - \frac{\chi_2}{V_\textrm{BC}} \frac{\kappa}{2 \pi} l^2.
\end{equation}
Since the computational result shows that the second term in the left hand side of Eq.\eqref{eq: Decay no-approx} is smaller than the first term by over an order of magnitude, we ignore the second term.
Then, Eq.\eqref{eq: Decay no-approx} has a simple solution of the form
\begin{equation} \label{eq: Simple Solution}
\frac{1}{l} = \frac{1}{l_0} + \frac{\chi_2}{V_\textrm{BC}} \frac{\kappa}{2 \pi} t,
\end{equation}
where $l_0$ is the initial TVLL at $t = t_\textrm{free}$.
Since $\mathcal{L} = l/V_\textrm{BC}$, the value of $\chi_2$ can be found as a slope by rescaling the vertical axis $1/l$ to be $(2 \pi / \kappa) (1/\mathcal{L})$.
Rescaling all the ``gradual decay'' and ``free decay'' profiles in Fig. \ref{fig: Point decay}(a), it turns out that all curves tends to collapse onto a single line of the slope $\chi_2 \approx 3.79$, as we can see in Fig. \ref{fig: Point decay}(b).
Considering the fact that the experimental value of $\chi_2$ in a homogeneous vortex tangle is of order unity, the value we obtained here with the localized vortex tangle may not be unreasonable, although the value of $\chi_2$ seems to be sensitive to the volume $V$ occupied by the vortex loops.
Here, we have applied the box-counting method to estimate the volume $V_\textrm{BC}$.
However, the way of estimating the occupied volume is not unique, and there remains some uncertainty in the determination of the value of $\chi_2$ in the localized vortex tangles.
Also, the approximation which allows us to obtain Eq.\eqref{eq: Simple Solution} only holds for sufficiently short time after turning off the heater, since the average distance among the vortices becomes larger than the average vortex loop size, which leads to the underestimation of the volume occupied by the vortices.
 The consequence of this effect can be observed in Fig. \ref{fig: Point decay}(b), \ie, the VLD $\mathcal{L}$ is overestimated, and the solid curves tend to drop below the broken curve whose slope is roughly $\chi_2 \approx 3.79$ after following it for at least a few second.

%
\section{Summary and Discussion}
We performed simulations based on VFM in order to investigate the evolution of a vortex tangle around a spherically symmetric thermal counterflow in superfluid \He.
For the spherically symmetric heat source, we considered two types of heaters.
One is a sphere with solid boundary on which vortices can connect, and the other is a point-like heater.
In both the cases, only the steady thermal counterflow profile is prescribed and any modulations in the normal-fluid profile
are not considered in this work.

Our simulations with a spherical heater reveal that a spherical vortex tangle is developed around it.
Under a strong radial counterflow, the orientation of vortex loops tends to be polarized and they increase in size.
When some vortices are connected to the solid wall of the heater, the tangle seems to grow unboundedly, forming a spiral-shape structure on vortex filaments.
The spiral structure may be identified as Kelvin waves that are excited on the filaments due to the thermal counterflow (DG instability).
The unbounded growth in vortex line would eventually lead our simulation to violate our assumption, namely, the disturbance in the normal fluid profile is no longer negligible.
In the vicinity of the heater, our current numerical scheme is, thus, only valid for simulating the early stages of the vortex tangle development, where radial vortex line density (RVLD) is $L \lesssim 10^5$ cm$^{-2}$.  
If the normal fluid is greatly blocked by a thick wall of vortex tangle of $L \gtrsim 10^8$ cm$^{-2}$, then a non-negligible temperature gradient may be accumulated within the tangle, as pointed out in the recent work by Sergeev \et \cite{Sergeev19}.
%

The simulations with a point-like heater resolve problem of the unbounded growth.
Since the radial counterflow drops as $1/r^2$, the vortex creation via DG instability becomes less dominant as vortices get farther away from the heat source.
Again, a spherical vortex tangle is formed around the point heater, but in this case, a hollow region is left inside the tangle, as the tangle front propagates radially outward.
The tangle development seems to have two phases; the production phase and the decay phase.
During the production phase, vortices are more or less confined in a small region of radius $R_\textrm{f}$.
Thus, the vortex loops in the tangle grow swiftly not only in lengths via mutual friction, but also in number because they repeatedly reconnect with each other within the region.
However, the growth is balanced by the decay due to the mutual friction, and the TVLL finds its maximum value.
During the decay period the number of small vortices in the tangle increases, and more vortices are able to ``escape'' the region of radius $R_\textrm{f}$, overcoming the counterflow.
Since the extra dissipation of vortex loop reduces the number of reconnections, the tangle starts to decay, which is what we call the ``gradual decay''.
If we turn off the point heater during the tangle development process, the tangle decays freely.
We call such a decay process the ``free decay'' in contrast to the the ``gradual decay.''
From the computational results we are able to extract the information of the decay constant $\chi_2$, although the value has an uncertainty since the volume that contains the vortices can be chosen somewhat arbitrarily.
%

In the present study, the normal-fluid profile is kept steady, however, such an assumption may not hold in the thermal boundary layer in the vicinity of the heater surface, or in a highly dense vortex tangle as we have discussed.
We would like to address these issues in the future work by coupling the normal-fluid dynamics to VFM.

M. T. acknowledges the support from JSPS KAKENHI (Grant No. JP17K05548, JP20H01855).

\bibliography{LocalHeatSpot_var2}

\end{document}